\documentclass[sigconf,nonacm]{acmart}

\usepackage{amsmath,bm}
\usepackage{booktabs,multirow,array,tabularx}
\usepackage{graphicx}
\usepackage{placeins}
\usepackage{needspace}
\usepackage{xurl}

\setcopyright{none}
\acmConference{}{}{}{}
\makeatletter
\renewcommand{\footnotetextcopyrightpermission}[1]{}
\makeatother

\newcommand{\AUC}{\ensuremath{\mathrm{AUC}}}
\newcommand{\MTR}{\ensuremath{\mathrm{MTR}}}

\newcommand{\FixedSpaceSection}[1]{\section{#1}}

\title{Inferring Hidden Structural Changes from Node-State Responses under Partial Network Observations}
\author{En Xu}
\affiliation{%
  \institution{Tsinghua University}
  \department{Department of Electronic Engineering}
  \city{Beijing}
  \country{China}}
\email{xuen@mail.tsinghua.edu.cn}
\author{Jingtao Ding}
\affiliation{%
  \institution{Tsinghua University}
  \department{Department of Electronic Engineering}
  \city{Beijing}
  \country{China}}
\email{dingjt15@tsinghua.org.cn}
\author{Yong Li}
\affiliation{%
  \institution{Tsinghua University}
  \department{Department of Electronic Engineering}
  \city{Beijing}
  \country{China}}
\email{liyong07@tsinghua.edu.cn}

\begin{document}

\begin{abstract}
Online platforms often expose activity signals even when records of the underlying connections are incomplete or delayed. These activity responses may retain indirect evidence of connections that formed during the observation period but are absent from the recorded topology. Recovering such hidden structural changes can help keep network records aligned with evolving interactions and support downstream network analysis.
However, the effects of a change propagate through the network and may overlap, making state changes difficult to associate with specific connections. We propose the Hierarchical Structural Context Ranker (HSCR), which combines visible network structure with node-state changes to infer connections that have formed but remain unrecorded. By jointly modeling the topological context of state responses and associations among potential connections, HSCR converts distributed activity changes into evidence for distinguishing potential connections without prespecified dynamical equations or parameters. Experiments on five temporal-network benchmarks and a real temporal Reddit network validate its effectiveness. Across observation spans and structural-record delays, true added connections achieve a 36\%--50\% reduction in mean true rank relative to the corresponding best baselines. This work provides a principled way to prioritize unrecorded additions for structural verification and record maintenance in online communities, information networks, and other networked systems.
\end{abstract}

\begingroup\emergencystretch=1em\maketitle\par\endgroup
\raggedbottom

\section{Introduction}
\label{sec:introduction}

\begingroup\emergencystretch=1em
Network representations describe complex systems through nodes and edges~\cite{newman2003structure,newman2010networks}. Temporal networks capture changing relationships~\cite{holme2012temporal,kazemi2020representation}, while response-based inference uses state observations to constrain hidden structure~\cite{timme2007revealing,ren2010inferring}. For Web interaction networks, we consider an observation gap: community activity summaries are available, but the visible connection record is incomplete or delayed. Can these summaries help rank connections that have already appeared but are absent from that record? This poses a graph-inference problem for understanding changes in online interactions under partial observation, with candidate rankings providing potential leads for structural verification.
\par\endgroup

\begingroup\emergencystretch=2em
Related research differs in both observations and targets. Static and temporal link prediction estimate potential future connections from topology, snapshots, or interaction events~\cite{liben2007link,lu2011link,kazemi2020representation}, whereas network reconstruction and dynamics learning infer latent relationships or state-transition models from trajectories~\cite{kipf2018neural}. Recent studies reconstruct network evolutionary histories from static network observations or augmented snapshots~\cite{wang2024reconstructing,xu2026diffusive}, demonstrating the value of recovering how network structure changes over time. Their focus is long-horizon historical reconstruction and depends on sufficiently informative temporal evidence. Our problem is different: given a partially observed graph and sparse node-state responses over a short interval, rank the candidate connections that account for the observed structural change rather than reconstructing the entire evolutionary history.
\par\endgroup

\begingroup\emergencystretch=3em
Networked systems often expose activity responses before the relations that generate them. Inferring latent or unrecorded associations from such responses can support structural verification and reveal mismatches between observed behavior and available records, providing a basis for anomaly screening when connections are missing, delayed, or changed unexpectedly. This capability also raises an information-leakage concern: behavioral responses may disclose sensitive relationships or attributes even when the underlying links are not directly observed, so inference should serve as audit evidence rather than proof of a causal structural event. This gap spans social, ecological, and brain networks, where activity can be measured while interactions, dependencies, or connectivity changes remain difficult to observe continuously~\cite{golder2014digital,lazer2009computational,bullmore2009complex,friston2011functional,kosinski2013private}.
\par\endgroup

\begin{figure*}[!t]
  \centering
  \includegraphics[width=0.85\linewidth]{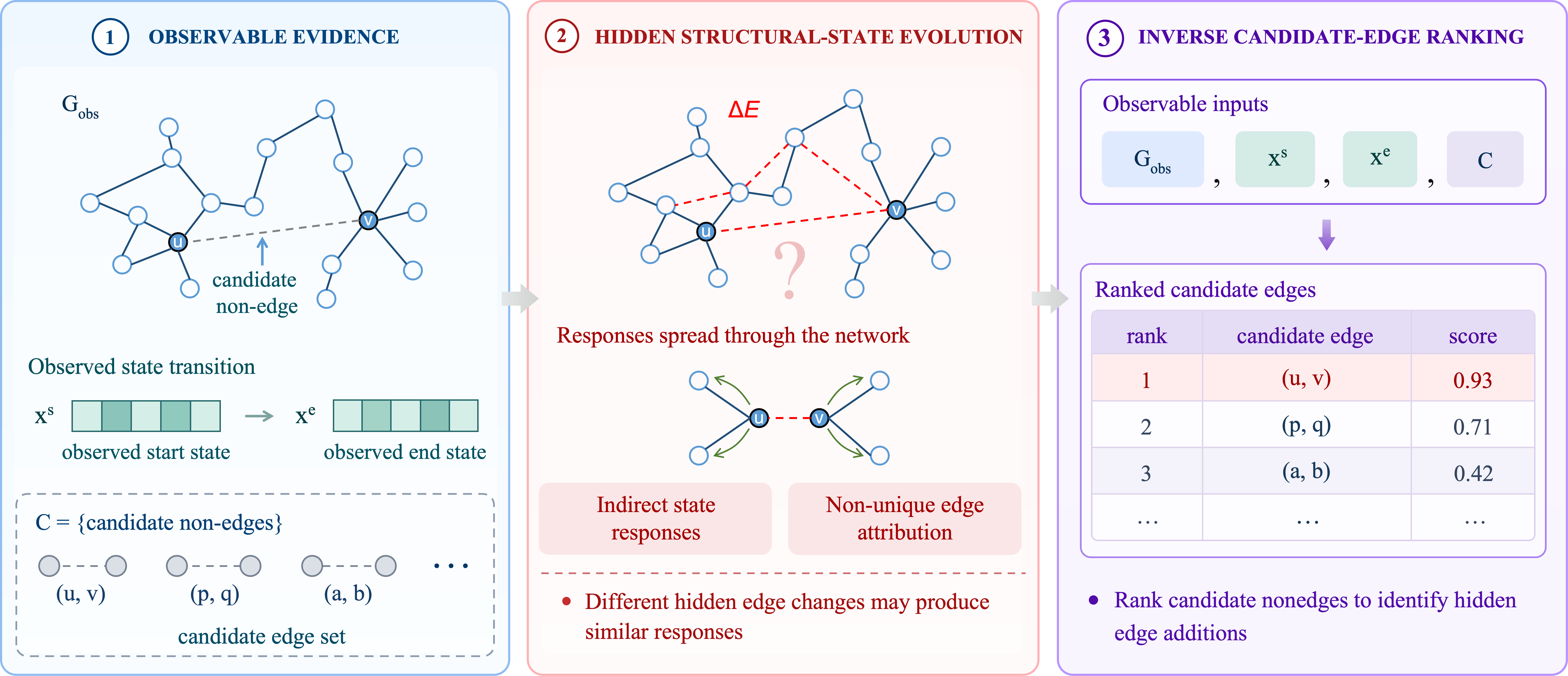}
  \caption{Inferring hidden edge additions from partial observations. Visible topology and endpoint states constrain the ranking of candidate connections whose formation is consistent with the observed state transition.}
  \label{fig:research-problem}
\end{figure*}

\begingroup\emergencystretch=1em
We study edge additions as an instance of hidden structural change. Given a visible topology, the node states at the two endpoints of an observation interval, and a domain of candidate nonedges, the task is to rank candidate connections that may have formed within the interval but are absent from the available structural record. The structural events and intermediate state trajectory inside the interval are not observed: new connections may form in one step or accumulate over several steps, and the visible snapshot may precede the state observations. The estimation target is therefore a conditional ranking of already occurred local changes under partial and temporally misaligned observations, rather than future link prediction, complete network reconstruction, or dynamical-system identification. Fig.~\ref{fig:research-problem} illustrates the observations and estimation target.
\par\endgroup

\begingroup\emergencystretch=2em
The central difficulty is that structural changes leave weak, indirect, and non-unique signatures in observed states. Network propagation can attenuate and diffuse a new connection's contribution, while unknown dynamics and sparse endpoint or snapshot observations prevent direct attribution. Effects from multiple changes may also superpose, allowing different candidate combinations to produce similar responses. The ranker must therefore attribute overlapping state evidence to specific candidates using partial structural context, without observing the dynamics or event sequence.
\par\endgroup

\begingroup\emergencystretch=2em
HSCR addresses this challenge by organizing scattered and potentially confounded state responses into candidate-attributable ranking evidence. It uses visible topology to establish the propagation context, anchors each response to candidate-centered local structure through endpoint-relative distances, roles, and neighborhood evidence, and reads candidate representations jointly over the complete candidate domain. This evidence-attribution chain links global propagation context, candidate-specific location, and shared or competing evidence among candidates, while requiring neither a prespecified dynamical model nor the unobserved event trajectory.
\par\endgroup

\begingroup\emergencystretch=1em
Across five temporal-network benchmarks and a real Reddit network, HSCR improves candidate discrimination and true-edge ranks across observation spans and structural-record delays. Ablations identify candidate-local structure and correct state--structure correspondence as important, with candidate-set context providing additional gains. Cross-network tests with simulated states show reusable but pair-dependent discriminative information, so transferability remains target-specific.
\par\endgroup

Our contributions are summarized as follows:
\begin{enumerate}
\begingroup\emergencystretch=1em
  \item We formulate hidden structural-change inference under partial and temporally misaligned observations as a distinct ranking problem, using node-state responses to prioritize unrecorded structural changes for structural verification and analysis of network evolution.
\par\endgroup
  \item We propose HSCR, which organizes state evidence through hierarchical structural context to rank hidden connection changes without prespecified dynamical equations or parameters.
  \item We evaluate HSCR on temporal-network benchmarks and Reddit, with ablations and cross-network tests probing its evidence sources and reuse beyond the training network.
\end{enumerate}

\section{Related Work}
\label{sec:related-work}

\subsection{Static Network Structure Modeling}

\begingroup\emergencystretch=2em
Static link-prediction methods derive candidate scores from observed topology. Local measures use neighborhood overlap, degree normalization, information rarity, or resource transfer (Common Neighbors, Jaccard, Adamic--Adar, and Resource Allocation)~\cite{liben2007link,lichtenwalter2010new,lu2011link,adamic2003friends,jaccard1901distribution,zhou2009predicting}; Katz and Local Path extend evidence to longer paths with different cost--coverage trade-offs~\cite{katz1953new,lu2009localpath,lu2011link}. Representation-learning methods encode co-occurrence or topology in node vectors (DeepWalk, node2vec, GAE/VGAE)~\cite{perozzi2014deepwalk,grover2016node2vec,kipf2016vgae}, while graph neural networks provide general neighborhood aggregation~\cite{scarselli2009graph,gilmer2017neural,kipf2016semi,velivckovic2018graph,xu2019powerful}. These node-level representations need not preserve candidate-specific distances, common-neigh\-borhood composition, or structural roles.
\par\endgroup

\begingroup\emergencystretch=2em
Candidate-centered methods instead organize evidence around a node pair. SEAL uses an enclosing subgraph and double-radius labeling~\cite{zhang2018seal}; NCN combines message passing with common-neigh\-bor features, and NCNC completes missing common-neigh\-bor structure with link predictions~\cite{wang2024ncnc}. We retain this candidate-local perspective but additionally use state responses before and after structural changes, restricting the target to connections that have already formed but remain unrecorded.
\par\endgroup

\subsection{Modeling Structural Evolution}

\begingroup\emergencystretch=2em
Dynamic-network research models changing relationships and their interaction with node states~\cite{zhang2015measuring,holme2012temporal,kazemi2020representation,sankar2020dysat}, including com\-plex-system predictability, physics-informed urban modeling, mobility characterization, and behavioral-attribute inference~\cite{xu2026predictability,xu2025survey,tan2021mobility,rostami2026personality}. Discrete-time methods use snapshots, whereas event-driven methods use interaction histories: DyRep models relationship and interaction processes~\cite{trivedi2019dyrep}, JODIE projects dynamic embeddings~\cite{kumar2019jodie}, and TGAT and Temporal Graph Networks combine temporal encoding, memory, and message passing~\cite{ma2020tgat,rossi2020temporal}. These approaches primarily support future-link, future-event, or state prediction from observed events.
\par\endgroup

Recent Web research also distinguishes forecasting activity from inferring its underlying network. Zhou et al.\ forecast node-state trajectories under unseen dynamical coefficients~\cite{zhou2026zeroshot}, whereas FIM infers diffusion networks from observed cascades~\cite{huang2024scalable}. Gan et al.\ jointly infer similar diffusion networks from limited diffusion observations~\cite{gan2026multitask}. Our target is instead local edge additions missing from a visible graph, ranked using interval-endpoint states within a specified candidate domain.

Latent-relation and task-driven structure-learning methods infer or select a graph for state prediction. Neural Relational Inference learns latent relations from trajectories~\cite{kipf2018neural}; discrete graph structure learning optimizes structure for a downstream task~\cite{franceschi2019learning}; and EvolveGCN adapts graph-convolution parameters to observed dynamic topologies~\cite{pareja2020evolvegcn}. Response-based inference and network tomography likewise use state changes to constrain unobserved relationships~\cite{timme2007revealing,ren2010inferring,steyvers2006inferring,castro2004network,newman2018network,deka2024learning}; related work also combines structural and steady-state information to infer formation order~\cite{xu2026inferring}. Unlike these methods, HSCR uses a possibly stale visible topology and only endpoint states to rank local additions within a candidate domain. It neither predicts future interactions nor recovers a complete network, event sequence, or state-transition law.

\section{Problem Formulation}
\label{sec:problem}

\subsection{Observations and Edge Additions}

\begingroup\emergencystretch=2em
Let $\{G_t\}_{t=0}^{T}$ be a sequence of discrete-time undirected simple graphs on a fixed node set $V$, where $G_t=(V,E_t)$ and $E_t\subseteq\binom{V}{2}$. The scalar state of node $i\in V$ at time $t$ is denoted by $x_{i,t}\in\mathbb R$, and the corresponding global state vector is $\bm x_t=[x_{1,t},\ldots,x_{|V|,t}]^\top\in\mathbb R^{|V|}$.
\par\endgroup

\begingroup\emergencystretch=2em
For each observation instance, let $G_{\mathrm{obs}}=(V,E_{\mathrm{obs}})$ denote the visible structure and $E_{\mathrm{tar}}$ the structural edge set at the target time. We assume that structural evolution within an observation instance follows a monotonic edge-ad\-dition process:
\par\endgroup
\begin{equation}
 E_{\mathrm{obs}}\subseteq E_{\mathrm{tar}},\qquad
 \Delta E=E_{\mathrm{tar}}\setminus E_{\mathrm{obs}},
 \label{eq:addition-process}
\end{equation}
where $\Delta E$ is the set of added edges not yet recorded in the visible structure. Equation~\eqref{eq:addition-process} defines the scope of estimation and excludes edge deletions. Each instance also provides an initial state $\bm x^s$ and a final state $\bm x^e$ for the interval, but not the structural evolution records within it.

\subsection{Candidate Domain and Ranking Objective}

Let
\begin{equation}
 \mathcal C=\{e_1,\ldots,e_m\}
 \subseteq \binom{V}{2}\setminus E_{\mathrm{obs}}
 \label{eq:candidate-domain}
\end{equation}
be the set of candidate nonedges for the current observation instance, where $m=|\mathcal C|$, satisfying the target-coverage condition $\Delta E\subseteq\mathcal C$. The information available for inference is represented by the observation tuple
\begin{equation}
 \mathcal O=\left(G_{\mathrm{obs}},\bm x^s,\bm x^e,\mathcal C\right).
 \label{eq:observation-tuple}
\end{equation}
The candidate construction rules are described in the experimental settings.

\begingroup\emergencystretch=2em
Inference uses only $\mathcal O$; dynamical equations and parameters, complete within-interval state trajectories, edge-level event times, candidate-edge provenance, and scenario identifiers are not provided to the ranker.
\par\endgroup

For a candidate edge $e_j\in\mathcal C$, define the membership label $y_j=\mathbb 1[e_j\in\Delta E]$ and let $\bm y=[y_1,\ldots,y_m]^\top\in\{0,1\}^{m}$. A model parameterized by $\theta$ generates a candidate score vector from $\mathcal O$:
\begin{equation}
 \bm s_\theta(\mathcal O)
 =f_\theta\!\left(G_{\mathrm{obs}},\bm x^s,\bm x^e,\mathcal C\right)
 =[s_1,\ldots,s_m]^\top\in\mathbb R^m,
 \label{eq:candidate-scores}
\end{equation}
Sorting these scores in descending order gives the candidate ranking:
\begin{equation}
 \pi_\theta
 =\operatorname*{argsort}_{j\in\{1,\ldots,m\}}(-s_j).
 \label{eq:candidate-order}
\end{equation}
The ranking objective is to favor candidates with $y_j=1$ over those with $y_k=0$. Evaluation uses the complete candidate domain $\mathcal C$, rather than only the subset used to compute the training loss. This objective is conditional on the candidate domain; $\mathcal C$ need not contain all nonedges of $G_{\mathrm{obs}}$. Different sets of added edges may produce identical or similar endpoint responses; the visible topology and candidate domain constrain the search but do not guarantee unique recovery.

\subsection{Three Observation Scenarios}

\begingroup\emergencystretch=2em
Real observations provide network snapshots and node states, but not the structural events between them. Added edges may form in one step or accumulate over a window, and structural records may lag behind state collection. We therefore consider three temporal alignments---synchronous, window-based, and stale-snapshot---under the common input form in \eqref{eq:observation-tuple}. The window span $w$ and structural lag $h$ specify temporal correspondence but are not inputs to candidate-edge scoring. A visual summary is provided in Appendix~\ref{app:observation-scenarios}.
\par\endgroup

\subsubsection{Scenario 1: Synchronous Edge-Addition Inference}

\begingroup\emergencystretch=2em
The most direct temporal alignment places the structural update and state transition within the same discrete step. Synchronous edge-ad\-dition inference uses the current structure and adjacent-time states to form the observation tuple
\par\endgroup
\begin{equation}
 \mathcal O_{\mathrm{syn}}
 =\left(G_t,\bm x_t,\bm x_{t+1},\mathcal C\right),\qquad
 \Delta E_{\mathrm{syn}}=E_{t+1}\setminus E_t.
 \label{eq:synchronous-addition}
\end{equation}
Under this controlled condition, all target edges form within one discrete step relative to the same initial structure, and ranking evidence over $\mathcal C$ comes from that step's state transition.

\subsubsection{Scenario 2: Window-Based Edge-Ad\-dition Lo\-calization}

\begingroup\emergencystretch=2em
When state measurements span multiple evolution steps and internal events are unobserved, the cumulative changes within the window must be identified from its initial structure and endpoint states. For an observation window of span $w$, the observation tuple and target set are
\par\endgroup
\begin{equation}
 \mathcal O_{\mathrm{win}}
 =\left(G_t,\bm x_t,\bm x_{t+w},\mathcal C\right),\qquad
 \Delta E_{\mathrm{win}}=E_{t+w}\setminus E_t.
\label{eq:window-addition}
\end{equation}
The target set contains the edges added cumulatively over $[t,t+w]$, which may have different effective durations of influence. The window-level membership label $\mathbb 1[e_j\in\Delta E_{\mathrm{win}}]$ encodes only membership in this added-edge set, not the insertion time.

\subsubsection{Scenario 3: Edge-Addition Inference from a Stale Snapshot}

When structure is collected less frequently than states, or structural records are delayed, the currently available structure may correspond to an earlier snapshot. More recent state responses must then be interpreted against an outdated structural context. When the structural observation lags behind the state observations by $h$ steps, the observation tuple and target set are
\begin{equation}
 \mathcal O_{\mathrm{stale}}
 =\left(G_{t-h},\bm x_t,\bm x_{t+1},\mathcal C\right),\;
 \Delta E_{\mathrm{stale}}=E_{t+1}\setminus E_{t-h}.
 \label{eq:stale-addition}
\end{equation}
This scenario uses an old structural snapshot to interpret near-current state responses. The target set covers all edges added after $G_{t-h}$ through $t+1$, rather than only those formed between adjacent state observations.

\section{HSCR: Hierarchical Structural Context Ranker}
\label{sec:method}

\subsection{Overview and Hierarchical Modeling}
\begin{figure*}[!t]
  \centering
  \includegraphics[width=\linewidth]{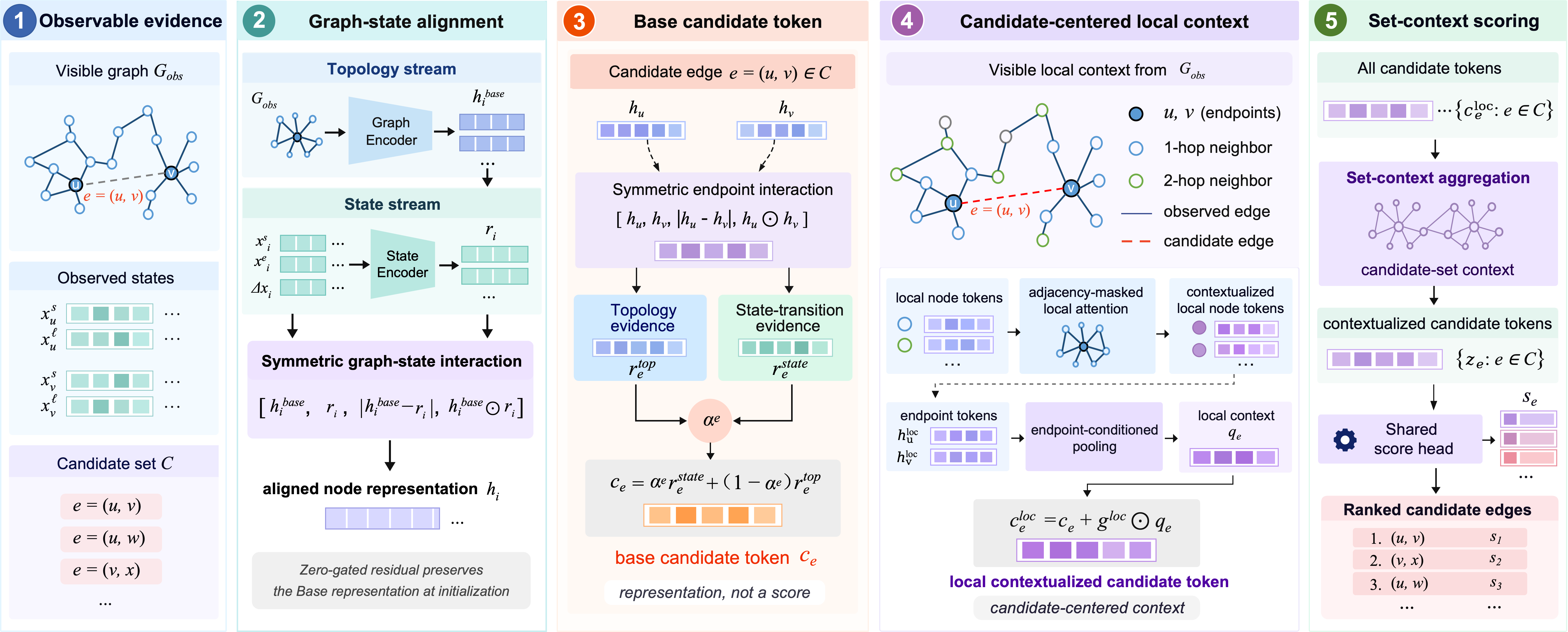}
  \caption{HSCR framework: visible-graph, candidate-local, and complete candidate-set contexts organize structural and state evidence for edge ranking. A shared scoring head produces candidate scores. Figure symbols $h_i^{\mathrm{base}}$, $q_e$, and $g^{\mathrm{loc}}$ correspond to $h_i^g$, $u_e^{\mathrm{loc}}$, and $\gamma_e^{\mathrm{loc}}$ in the text; see \eqref{eq:hscr-local-update} for the local update.}
  \label{fig:hscr-framework}
\end{figure*}

\begingroup\emergencystretch=2em
Given the observation tuple $\mathcal O=(G_{\mathrm{obs}},\bm x^s,\bm x^e,\mathcal C)$ defined in Section~\ref{sec:problem}, HSCR ranks candidate edges by organizing evidence at three levels: visible-graph context, candidate-centered local structure, and relationships within the complete candidate set. Its representation flow is
\par\endgroup
\begin{equation}
 \mathcal O\longrightarrow h_i^{g}\longrightarrow h_i\longrightarrow c_e\longrightarrow c_e^{\mathrm{loc}}\longrightarrow z_e\longrightarrow s_e .
 \label{eq:hscr-flow}
\end{equation}
\begingroup\emergencystretch=2em
Here $h_i^g$ is a visible-graph representation, $h_i$ is state-conditioned, $c_e$ is a base candidate representation, and the last three terms add local context, set context, and a scalar ranking score. Candidate nonedges never enter message passing on $G_{\mathrm{obs}}$; refinement modules use zero-initialized gates. Fig.~\ref{fig:hscr-framework} summarizes this evidence-attribution chain.
\par\endgroup

\subsection{Evidence and Base Candidate Encoding}
For node $i$, let $d_i$ be its visible-graph degree and $\Delta x_i=x_i^e-x_i^s$. The node evidence is
\begin{equation}
 a_i=[x_i^s,x_i^e,\Delta x_i,d_i]\in\mathbb R^4 .
 \label{eq:hscr-node-evidence}
\end{equation}
\begingroup\emergencystretch=2em
For a canonicalized undirected candidate $e=(u_e,v_e)$, endpoint exchange consistency is enforced with $u_e=\min(u,v)$ and $v_e=\max(u,v)$. For each endpoint-wise state quantity $p\in\{x^s,\allowbreak x^e,\allowbreak\Delta x,\allowbreak|\Delta x|\}$, we use
\par\endgroup
\begin{equation}
 \rho(p_u,p_v)=[p_u+p_v,\ |p_u-p_v|,p_up_v].
 \label{eq:hscr-symmetric-lift}
\end{equation}
Together with symmetric topology statistics this gives
\begin{equation}
 f_e=[f_e^{\mathrm{top}},f_e^{\mathrm{state}}]\in\mathbb R^{23},\quad f_e^{\mathrm{top}}\in\mathbb R^9,\quad f_e^{\mathrm{state}}\in\mathbb R^{14};
 \label{eq:hscr-edge-evidence}
\end{equation}
\begingroup\emergencystretch=2em
the complete feature list is given in Appendix~\ref{app:evidence}. The node encoder applies mean message passing only on $E_{\mathrm{obs}}$ and a state-conditioned gated refinement, producing $h_i^g$ and $h_i$. For a candidate, endpoint interactions feed topology and state branches; a scalar gate combines them as
\par\endgroup
\begingroup
\medmuskip=2mu
\thickmuskip=3mu
\begin{equation}
 c_e=\alpha_e r_e^{\mathrm{state}}+(1-\alpha_e)r_e^{\mathrm{top}},\; \alpha_e=\sigma(\operatorname{Gate}_{\mathrm{edge}}([r_e^{\mathrm{state}},r_e^{\mathrm{top}}])).
 \label{eq:hscr-base-edge-token}
\end{equation}
\endgroup
The message-passing and branch equations are provided in Appendix~\ref{app:evidence}.

\subsection{Candidate-Centered Local Context}
Whole-graph propagation supplies shared context but does not identify where a response lies relative to one candidate. For $e=(u_e,v_e)$, HSCR extracts
\begin{equation}
 \mathcal V_e^{\mathrm{loc}}=\{j\in V:\min(\operatorname{dist}(u_e,j),\operatorname{dist}(v_e,j))\le r\},
 \label{eq:hscr-local-node-set}
\end{equation}
retaining both endpoints and only visible adjacency; the candidate edge itself is never inserted. Local node representations encode endpoint-relative distances and structural roles. After masked local propagation and endpoint-centered pooling, the increment updates the base candidate representation:
\begin{equation}
 c_e^{\mathrm{loc}}=c_e+\gamma_e^{\mathrm{loc}}\odot u_e^{\mathrm{loc}},\qquad \gamma_e^{\mathrm{loc}}=\tanh(\operatorname{Gate}_{\mathrm{loc}}([c_e,u_e^{\mathrm{loc}}])).
 \label{eq:hscr-local-update}
\end{equation}
The complete local token, deterministic truncation rule, masked attention, and pooling equations appear in Appendix~\ref{app:local}.

\subsection{Complete Candidate-Set Context}
Local representations are candidate-specific. To expose relations among all $m=|\mathcal C|$ candidates without an $m\times m$ attention matrix, HSCR forms $K$ inducing summaries. With $q_k=W_QI_k$, $k_e=W_Kc_e^{\mathrm{loc}}$, and $v_e=W_Vc_e^{\mathrm{loc}}$,
\begin{equation}
 \begin{aligned}
 w_{ke}&=\sigma(q_k^\top k_e/\sqrt d), & H_k&=N_k/\max(D_k,\epsilon),\\
 N_k&=\sum_{e\in\mathcal C}w_{ke}v_e, & D_k&=\sum_{e\in\mathcal C}w_{ke}.
 \end{aligned}
 \label{eq:hscr-set-summary}
\end{equation}
Each candidate reads the shared summaries and receives a gated update, $z_e=c_e^{\mathrm{loc}}+\gamma_e^{\mathrm{set}}\odot\delta_e^{\mathrm{set}}$ (full readback in Appendix~\ref{app:set}). Additivity permits blockwise computation and gives $O(mK)$ set interaction rather than pairwise attention; negative sampling affects only the loss, not the complete set context.

\subsection{Ranking Objective and Training}
A shared head produces
\begin{equation}
 s_e=\operatorname{ScoreHead}(z_e),\qquad e\in\mathcal C,
 \label{eq:hscr-score}
\end{equation}
with scores ordered as in Section~\ref{sec:problem}. HSCR jointly optimizes pairwise ranking, a set-level margin, and an auxiliary count prediction:
\begin{equation}
 \mathcal L=\mathcal L_{\mathrm{pair}}+\lambda_{\mathrm{set}}\mathcal L_{\mathrm{set}}+\lambda_{\mathrm{count}}\mathcal L_{\mathrm{count}}.
 \label{eq:hscr-total-loss}
\end{equation}
These terms constrain individual orderings, the aggregate positive--negative gap, and the number of positives; the count head is not used for scoring. Their full definitions and the two-stage optimization procedure are given in Appendix~\ref{app:training}. Zero-initialized gates provide $h_i=h_i^g$, $c_e^{\mathrm{loc}}=c_e$, and $z_e=c_e$ at the start of the second stage.

\section{Experiments and Results}

\subsection{Experimental Settings}
\begingroup\emergencystretch=2em
We evaluate HSCR on controlled networks and Reddit, then analyze evidence sources, dynamics, observation quality, scale, training-data volume, and cross-network transfer.
\par\endgroup

\subsubsection{Datasets and State Observations}

\paragraph{Controlled temporal network topologies}
\begingroup\emergencystretch=2em
The controlled experiments use five timestamped edge-event datasets—\textit{Chaos} (CHA), \textit{Complex Networks} (COM), \textit{Fluctuations} (FLU), \textit{Interfaces} (INT), and \textit{Thermodynamics} (THE)~\cite{xu2026diffusive}. These labels do not denote specific real-world domains. From each sequence of undirected events we construct cumulative snapshots, treating repeated events as one existing edge and including every observed node. Table~\ref{tab:controlled_data_scale} summarizes their sizes, additions, and densities.

\par\endgroup

\begin{table}[!t]
  \centering
  \caption{Controlled dataset statistics: nodes ($N$), final cumulative unique edges ($M$), time points ($T$), and per-step edge additions.}
  \label{tab:controlled_data_scale}
  \small
  \setlength{\tabcolsep}{2pt}
  \begin{tabular*}{0.95\columnwidth}{@{\extracolsep{\fill}}lrrrrrr@{}}
    \toprule
    \multirow{2}{*}[-2pt]{Dataset} & \multirow{2}{*}[-2pt]{$N$} & \multirow{2}{*}[-2pt]{$M$} & \multirow{2}{*}[-2pt]{$T$} & \multicolumn{2}{c}{Edges added} & \multirow{2}{*}[-2pt]{Density} \\
    \cmidrule(lr){5-6}
    & & & & Mean & Max. & \\
    \midrule
    \textit{Chaos} & 2055 & 3758 & 1118 & 3.361 & 39 & 0.001781 \\
    \textit{Complex Networks} & 225 & 413 & 172 & 2.401 & 12 & 0.016389 \\
    \textit{Fluctuations} & 1248 & 2198 & 731 & 3.007 & 45 & 0.002825 \\
    \textit{Interfaces} & 2745 & 6718 & 1040 & 6.460 & 109 & 0.001784 \\
    \textit{Thermodynamics} & 158 & 228 & 131 & 1.740 & 8 & 0.018383 \\
    \bottomrule
  \end{tabular*}
\end{table}

\paragraph{Simulated dynamics and state-observation generation}
\begingroup\emergencystretch=2em
Initial states are sampled uniformly from $[0,1]$ and advanced with the corresponding rule, using the intermediate snapshots for window observations and first advancing to the state-observation time for stale snapshots. Each update combines a node's state with the mean neighbor state; configured Gaussian noise is then added and states are clipped to their domain. Standard scenarios use DeGroot dynamics with $\alpha=0.5$, while cross-dynamics tests use four alternative rules. Appendix~\ref{app:dynamics} provides the full update equations and default parameters.
\par\endgroup

\paragraph{Reddit dataset with real observations}
The real-observation experiments use SNAP subreddit hyperlink title events~\cite{leskovec2014snap,golder2014digital,lazer2009computational}. Subreddits are nodes, hyperlink pairs are canonicalized as undirected edges, and monthly aggregation yields structural snapshots; node states aggregate activity and sentiment. The construction of this derived Reddit benchmark, including filtering, state formation, candidate generation, and temporal splitting, is detailed in Appendix~\ref{app:reddit-data}.

\subsubsection{Comparison Methods}
Comparisons cover nonlearned scoring, explicit-feature learning, and structural representation learning. Every method is adapted to the same candidate-edge scoring interface.

\begingroup\emergencystretch=2em
Nonlearned baselines use either topology or endpoint-state changes: Common Neighbors (CN), Jaccard Coefficient (JC), Adamic--Adar (AA), and Resource Allocation (RA) capture neighborhood overlap, degree normalization, information rarity, and local resource transfer~\cite{liben2007link,adamic2003friends,jaccard1901distribution,zhou2009predicting}, whereas State Difference (SD) ranks candidates by endpoint-state changes. These methods have no trainable parameters. SSFS (Structure--State Feature Scorer) is our explicit-feature MLP baseline~\cite{rumelhart1986learning}; it combines local topological and endpoint-state features without visible-graph message passing, candidate-centered context, or candidate-set context.

Network-oriented baselines cover graph representation learning, candidate-centered subgraph reasoning, and latent-relation inference. NCNC combines visible-graph representations with common-neighbor completion~\cite{wang2024ncnc}; NRI infers candidate relations from endpoint states, and GTS models temporal state relationships from two frames~\cite{kipf2018neural,shang2021discrete}. PENCIL uses a fixed-size candidate-centered enclosing subgraph~\cite{zhang2018seal,teru2020inductive}; GRN combines historical graph context with current endpoint states~\cite{trivedi2019dyrep,rossi2020temporal,pareja2020evolvegcn}; LLACE uses locally linear contrastive graph representations, while TER-AER uses edge-wise observation representations~\cite{liu2024llace,wang2023efficient}. All methods are adapted to the same candidate-edge scoring interface; comparisons preserve their modeling emphasis and do not claim full reproduction under other tasks or inputs.

\par\endgroup

\begin{figure}[!t]
  \setlength{\abovecaptionskip}{5pt}
  \centering
  \includegraphics[width=0.9\columnwidth]{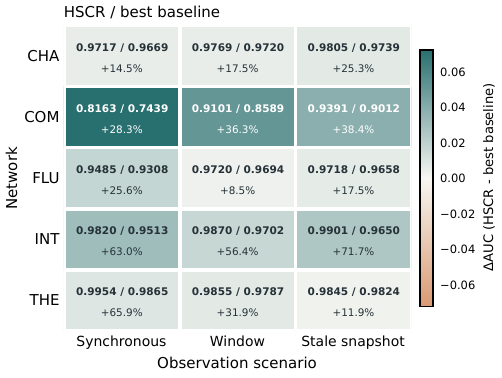}
  \caption{AUC and normalized gains across five temporal networks and three observation scenarios.}
  \label{fig:controlled-heatmap}
\end{figure}

\begin{figure}[!t]
  \setlength{\abovecaptionskip}{1pt}
  \centering
  \hspace*{-0.02\columnwidth}\includegraphics[width=0.81\columnwidth]{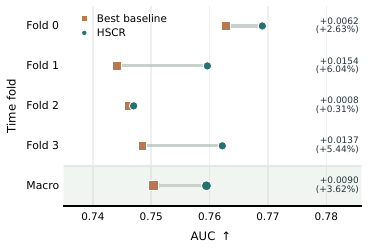}
  \caption{Reddit AUC under existing-link activity states: HSCR versus the best baseline across time folds. Annotations show absolute and normalized AUC gains.}
  \label{fig:reddit-repeat}
\end{figure}

\begin{figure*}[!t]
  \setlength{\abovecaptionskip}{1pt}
  \centering
  \includegraphics[width=0.7885\linewidth,trim=0bp 3.37bp 0bp 0bp,clip]{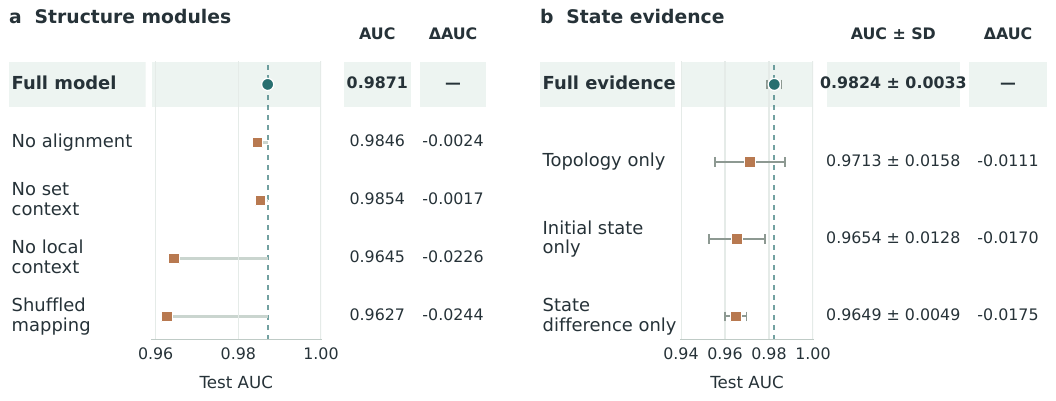}
  \caption{Ablations on Interfaces. (a) Structural modules under window localization; points show AUC. (b) State evidence under synchronous inference; points and error bars show mean $\pm$ SD over three training seeds.}
  \label{fig:mechanism-analysis}
\end{figure*}

\subsection{Overall Ranking Performance}
\expandafter\def\csname BaselineexpHSCRauc\endcsname{0.9390}
\expandafter\def\csname BaselineexpHSCRmtr\endcsname{955.7}
\expandafter\def\csname BaselineexpSamerunBaseauc\endcsname{0.9334}
\expandafter\def\csname BaselineexpSamerunBasemtr\endcsname{1441.3}
\expandafter\def\csname BaselineexpRandomauc\endcsname{0.4823}
\expandafter\def\csname BaselineexpRandommtr\endcsname{15943.9}
\expandafter\def\csname BaselineexpStateheuristicauc\endcsname{0.4240}
\expandafter\def\csname BaselineexpStateheuristicmtr\endcsname{17230.3}
\expandafter\def\csname BaselineexpHSCRauc\endcsname{0.9636}
\expandafter\def\csname BaselineexpHSCRmtr\endcsname{715.5}
\expandafter\def\csname BaselineexpSamerunBaseauc\endcsname{0.9532}
\expandafter\def\csname BaselineexpSamerunBasemtr\endcsname{1383.0}
\expandafter\def\csname BaselineexpNoFCRGauc\endcsname{0.9594}
\expandafter\def\csname BaselineexpNoFCRGmtr\endcsname{1248.0}
\expandafter\def\csname BaselineexpRandomauc\endcsname{0.5122}
\expandafter\def\csname BaselineexpRandommtr\endcsname{16060.6}
\expandafter\def\csname BaselineexpStateheuristicauc\endcsname{0.5084}
\expandafter\def\csname BaselineexpStateheuristicmtr\endcsname{16246.4}
\expandafter\def\csname BaselineexpHSCRauc\endcsname{0.9749}
\expandafter\def\csname BaselineexpHSCRmtr\endcsname{516.8}
\expandafter\def\csname BaselineexpSamerunBaseauc\endcsname{0.9676}
\expandafter\def\csname BaselineexpSamerunBasemtr\endcsname{1035.1}
\expandafter\def\csname BaselineexpNoFCRGauc\endcsname{0.9629}
\expandafter\def\csname BaselineexpNoFCRGmtr\endcsname{1026.2}
\expandafter\def\csname BaselineexpRandomauc\endcsname{0.4908}
\expandafter\def\csname BaselineexpRandommtr\endcsname{16013.9}
\expandafter\def\csname BaselineexpStateheuristicauc\endcsname{0.6543}
\expandafter\def\csname BaselineexpStateheuristicmtr\endcsname{10505.9}
\expandafter\def\csname BaselineredditfullHSCRauc\endcsname{0.8444}
\expandafter\def\csname BaselineredditfullHSCRmtr\endcsname{4913.9}
\expandafter\def\csname BaselineredditfullSamerunBaseauc\endcsname{0.8370}
\expandafter\def\csname BaselineredditfullSamerunBasemtr\endcsname{5156.2}
\expandafter\def\csname BaselineredditfullNoFCRGauc\endcsname{0.8317}
\expandafter\def\csname BaselineredditfullNoFCRGmtr\endcsname{5317.5}
\expandafter\def\csname BaselineredditfullRandomauc\endcsname{0.5022}
\expandafter\def\csname BaselineredditfullRandommtr\endcsname{15476.5}
\expandafter\def\csname BaselineredditfullStateheuristicauc\endcsname{0.4426}
\expandafter\def\csname BaselineredditfullStateheuristicmtr\endcsname{17377.3}
\expandafter\def\csname BaselineredditrepeatHSCRauc\endcsname{0.7646}
\expandafter\def\csname BaselineredditrepeatHSCRmtr\endcsname{7351.8}
\expandafter\def\csname BaselineredditrepeatSamerunBaseauc\endcsname{0.7496}
\expandafter\def\csname BaselineredditrepeatSamerunBasemtr\endcsname{7812.0}
\expandafter\def\csname BaselineredditrepeatNoFCRGauc\endcsname{0.7496}
\expandafter\def\csname BaselineredditrepeatNoFCRGmtr\endcsname{7812.0}
\expandafter\def\csname BaselineredditrepeatRandomauc\endcsname{0.5022}
\expandafter\def\csname BaselineredditrepeatRandommtr\endcsname{15476.5}
\expandafter\def\csname BaselineredditrepeatStateheuristicauc\endcsname{0.4610}
\expandafter\def\csname BaselineredditrepeatStateheuristicmtr\endcsname{16785.9}

We compare all methods on the same observation instances and complete candidate sets across the three controlled tasks. AUC and mean true rank (MTR) measure candidate discrimination and true-edge positions (higher AUC and lower MTR are better); AUC is threshold-independent~\cite{hanley1982meaning}. Table~\ref{tab:controlled-overall} reports task-level macro averages.

We define the normalized AUC gain as
\[
 \mathrm{Gain}_{\mathrm{AUC}} = \frac{\mathrm{AUC}_{\mathrm{HSCR}}-\mathrm{AUC}_{\mathrm{best\ baseline}}}{1-\mathrm{AUC}_{\mathrm{best\ baseline}}},
\]
the fraction of the best baseline's remaining error eliminated by HSCR; for MTR, the relative reduction is $(b-h)/b$, where $b$ and $h$ denote the best-baseline and HSCR MTRs, respectively, so positive values indicate lower mean true rank.

\begin{table*}[!t]
  \centering
  \caption{Overall results on the three controlled observation tasks; differences are computed relative to the best baseline, and MTR reduction is normalized by the baseline MTR.}
  \label{tab:controlled-overall}
  \small
  \setlength{\tabcolsep}{3pt}
  \begin{tabular}{lcccccccc}
    \toprule
    Scenario & \begin{tabular}[c]{@{}c@{}}HSCR\\AUC\end{tabular} & \begin{tabular}[c]{@{}c@{}}Best baseline\\AUC\end{tabular} & $\Delta$AUC & AUC gain & \begin{tabular}[c]{@{}c@{}}HSCR\\MTR\end{tabular} & \begin{tabular}[c]{@{}c@{}}Best baseline\\MTR\end{tabular} & $\Delta$MTR & \begin{tabular}[c]{@{}c@{}}MTR\\reduction\end{tabular} \\
    \midrule
    Synchronous & \textbf{0.9428} & 0.9159 & +0.0269 & 31.96\% & \textbf{974.6} & 1693.1 & -718.5 & 42.43\% \\
    Window & \textbf{0.9663} & 0.9498 & +0.0164 & 32.78\% & \textbf{689.3} & 1082.5 & -393.1 & 36.32\% \\
    Stale snapshot & \textbf{0.9732} & 0.9577 & +0.0156 & 36.76\% & \textbf{534.5} & 1073.0 & -538.5 & 50.19\% \\
    \bottomrule
  \end{tabular}
\end{table*}

Table~\ref{tab:controlled-overall} shows HSCR AUCs of $0.9428$, $0.9663$, and $0.9732$, with absolute improvements of $0.0269$, $0.0164$, and $0.0156$ and normalized gains of $31.96\%$, $32.78\%$, and $36.76\%$ over the corresponding baselines' remaining error. MTR decreases by $718.5$, $393.1$, and $538.5$, corresponding to relative reductions of $42.43\%$, $36.32\%$, and $50.19\%$; Fig.~\ref{fig:controlled-heatmap} gives network-level comparisons.

\begin{figure}[!t]
  \setlength{\abovecaptionskip}{1pt}
  \centering
  \includegraphics[width=0.98\columnwidth,trim=0bp 3.37bp 0bp 0bp,clip]{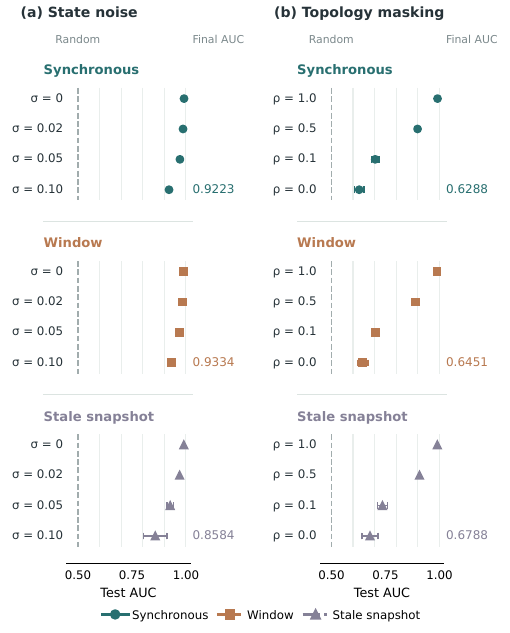}
  \caption{Test AUC under state noise (a) and visible-topology masking (b). Horizontal error bars show the sample standard deviation across three training seeds; the vertical dashed line denotes random ranking.}
  \label{fig:observation-robustness}
\end{figure}

\begin{figure}[!t]
  \setlength{\abovecaptionskip}{1pt}
  \centering
  \includegraphics[width=0.98\columnwidth,trim=0bp 1.70bp 0bp 0bp,clip]{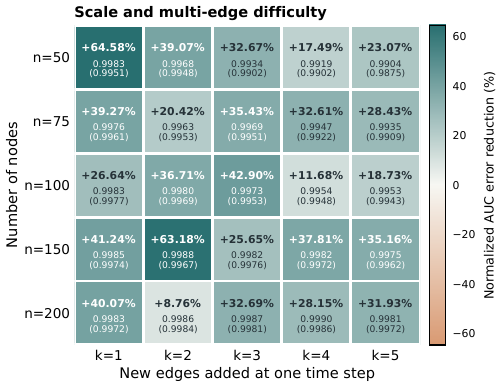}
  \caption{AUC and normalized gains across network sizes and simultaneous edge additions.}
  \label{fig:scale-multiedge}
\end{figure}

\subsection{External Validation on Reddit}

\begingroup\emergencystretch=2em
To assess applicability to real observations, we evaluate HSCR on identifying community-pair hyperlinks that first appear in a temporal Reddit record. The benchmark uses monthly aggregation, with construction details given in Appendix~\ref{app:reddit-data}. Because the same events define both structural additions and node-state statistics, we compare two state views: full states include all retained events, whereas existing-link activity states exclude, from state aggregation only, events on edges first appearing in the target month. Structural snapshots, candidate sets, and target labels are shared across the two views, isolating whether ranking can rely on existing structure and activity rather than the direct contribution of target-month events.
\par\endgroup
\begin{table}[!t]
\centering\small
\setlength{\tabcolsep}{2.5pt}
\caption{Reddit comparisons under two state conditions. Bold marks HSCR; underlining marks the best baseline. Parentheses give normalized AUC gain, relative MTR reduction, and relative Recall@100 increase.}
\label{tab:reddit-comparison}
\begin{tabular}{lcccccc}
\toprule
\multirow{2}{*}[-6pt]{Method} & \multicolumn{3}{c}{Full states} & \multicolumn{3}{c}{Existing-link activity states} \\
\cmidrule(lr){2-4}\cmidrule(lr){5-7}
 & AUC $\uparrow$ & MTR $\downarrow$ & \begin{tabular}[c]{@{}c@{}}Recall@\\100 $\uparrow$\end{tabular} & AUC $\uparrow$ & MTR $\downarrow$ & \begin{tabular}[c]{@{}c@{}}Recall@\\100 $\uparrow$\end{tabular} \\
\midrule
\multicolumn{7}{l}{\textit{Heuristic baselines}} \\
SD & 0.4426 & 17377.3 & 0.0022 & 0.4610 & 16785.9 & 0.0034 \\
CN & 0.6328 & 11325.6 & 0.0082 & 0.6328 & 11325.6 & 0.0082 \\
JC & 0.5812 & 12993.6 & 0.0054 & 0.5812 & 12993.6 & 0.0054 \\
AA & 0.6358 & 11284.3 & 0.0071 & 0.6358 & 11284.3 & 0.0071 \\
RA & 0.6469 & 10950.4 & 0.0061 & 0.6469 & 10950.4 & 0.0061 \\
\addlinespace[2pt]
\multicolumn{7}{l}{\textit{Learning-based baselines}} \\
SSFS & 0.7826 & 6844.6 & 0.0194 & 0.6875 & 9746.7 & 0.0207 \\
LLACE & 0.7977 & 5987.8 & 0.0482 & 0.6855 & 9233.6 & 0.0200 \\
TER-AER & 0.8001 & 6305.9 & 0.0408 & 0.6961 & 9485.9 & 0.0150 \\
NCNC & 0.8327 & 5284.3 & \underline{0.0622} & \underline{0.7510} & 7782.5 & 0.0379 \\
NRI & 0.8325 & 5290.6 & 0.0600 & 0.7468 & 7916.3 & 0.0362 \\
GTS & 0.8318 & 5313.6 & 0.0576 & 0.7461 & 7938.6 & 0.0335 \\
PENCIL & \underline{0.8336} & \underline{5251.0} & 0.0553 & 0.7477 & 7870.3 & 0.0310 \\
GRN & 0.8280 & 5260.7 & 0.0572 & 0.7338 & \underline{7772.3} & \underline{0.0384} \\
\addlinespace[2pt]
\multicolumn{7}{l}{\textit{Our method}} \\
HSCR & \begin{tabular}[c]{@{}c@{}}\textbf{0.8444} \\ (+6.47\%)\end{tabular} & \begin{tabular}[c]{@{}c@{}}\textbf{4913.9} \\ (+6.42\%)\end{tabular} & \begin{tabular}[c]{@{}c@{}}\textbf{0.0770} \\ (+23.70\%)\end{tabular} & \begin{tabular}[c]{@{}c@{}}\textbf{0.7646} \\ (+5.46\%)\end{tabular} & \begin{tabular}[c]{@{}c@{}}\textbf{7351.8} \\ (+5.41\%)\end{tabular} & \begin{tabular}[c]{@{}c@{}}\textbf{0.0392} \\ (+2.02\%)\end{tabular} \\
\bottomrule
\end{tabular}
\end{table}

Table~\ref{tab:reddit-comparison} reports HSCR AUCs of $0.8444$ and $0.7646$ under the two conditions, above the best baselines ($0.8336$ PENCIL and $0.7510$ NCNC), with MTRs of $4913.9$/$5251.0$ and $7351.8$/$7782.5$. Topological heuristics beat random ranking, whereas direct state differences fall below random; this reflects the weakness of that score, not of state information itself. Recall@100 is $0.0770$/$0.0392$, so top-ranked coverage remains limited.

\begingroup\setlength{\parskip}{0pt}
Removing first-appearance events from state statistics lowers all learned methods, including HSCR (from $0.8444$ to $0.7646$), but HSCR retains candidate discrimination. The remaining information may come from visible topology, existing-link activity, and their joint use; Fig.~\ref{fig:reddit-repeat} compares the resulting AUC.
\par\endgroup

\subsection{Analysis of Model Components}
\begingroup\emergencystretch=2em
We conduct two ablation groups on Interfaces. Structural-module ablations use window localization and one training seed; state-evidence ablations use synchronous inference and report \mbox{mean$\pm$SD} over three seeds. Because protocols differ, comparisons are within groups only.
\par\endgroup

In Fig.~\ref{fig:mechanism-analysis}(a), the full model reaches AUC $0.9871$. Removing local candidate context causes the largest drop ($0.0226$); removing candidate-set context or state--structure alignment reduces AUC by $0.0017$ or $0.0024$. Shuffling the state--structure correspondence reduces AUC by $0.0244$, confirming that ranking depends on matching state evidence to candidate-specific structural positions.

Fig.~\ref{fig:mechanism-analysis}(b) shows that full state evidence yields $0.9824\pm0.0033$, improving by $0.0111$, $0.0170$, and $0.0175$ over topology-only, initial-state-only, and state-difference-only variants, respectively.

\subsection{Dynamics and Observation Robustness}

We examine robustness to state-generation mechanisms and observation quality. Cross-dynamics experiments use nonlinear response, threshold propagation, SIS mean-field, and coupled logistic dynamics across the three scenarios, yielding 12 cells over five networks and three seeds. Models are trained independently; neither HSCR nor baselines receive the dynamical type, equation, or parameters.

Table~\ref{tab:cross-dynamics} reports normalized gains over the strongest baseline's remaining error. All 12 cells are positive ($21.94\%$--$46.29\%$) with corresponding MTR improvements. Because each dynamics condition is trained separately, this supports applicability across rules, not zero-shot transfer.

\begin{table}[!b]
  \centering
  \caption{Macro-average AUC gains over the strongest baseline under four alternative dynamics. Gains use the baseline's remaining-error normalization.}
  \label{tab:cross-dynamics}
  \small
  \setlength{\tabcolsep}{5pt}
  \begin{tabular}{lrrrr}
    \toprule
    \multirow{2}{*}[-2pt]{Scenario} & \multicolumn{4}{c}{AUC gain over the strongest baseline} \\
    \cmidrule(lr){2-5}
    & Nonlinear & Threshold & SIS & Logistic \\
    \midrule
    Synchronous & 21.94\% & 38.08\% & 34.52\% & 29.94\% \\
    Window & 29.95\% & 42.49\% & 40.46\% & 41.42\% \\
    Stale snapshot & 34.23\% & 38.13\% & 46.29\% & 38.09\% \\
    \bottomrule
  \end{tabular}
\end{table}

Observation-perturbation experiments instead assess degradation under noisy states or incomplete structure. All scenarios use Interfaces, DeGroot ($\alpha=0.5$), identical node sets, candidate construction, splits, seeds, and evaluation, differing only in synchronous, window-accumulated, or stale-snapshot alignment. At test time we vary $\sigma\in\{0,0.02,0.05,0.10\}$ or $\rho\in\{1.0,0.5,0.1,0.0\}$. Noise is absent from training, validation, and clean test states, and is added independently to both test endpoints without clipping. Repeats are averaged within each model, then means and sample SDs are computed across three seeds; deterministic conditions are evaluated once per model. Fig.~\ref{fig:observation-robustness} reports AUC, with $0.5$ denoting random ranking.

All scenarios degrade as state noise increases; at $\sigma=0.10$, AUC is $0.9223$, $0.9334$, and $0.8584$ for synchronous, window, and stale-snapshot settings. Topology masking also degrades performance overall; at $\rho=0$, the corresponding AUCs are $0.6288$, $0.6451$, and $0.6788$. Under the common-network protocol, sensitivity differs across temporal conditions: stale snapshots decline more at the highest noise but retain higher mean AUC at the tested masking levels.

\subsection{Network Size and Multiple Edge Additions}

We vary whole-graph size (50, 75, 100, 150, 200 nodes) and simultaneous additions (1--5) in 25 synchronous cells, each with 32 networks, 18 time steps, and three seeds. Data generation, candidates, and training are fixed; the strongest baseline is selected on validation. The local encoder caps its processed node count at 64, independent of graph size. Fig.~\ref{fig:scale-multiedge} reports mean AUC and normalized gains.

\begingroup\emergencystretch=2em
Across the 25 size--addition combinations, HSCR achieves a macro-average AUC of $0.9967$, exceeding the strongest baseline's $0.9952$. The mean normalized gain is $32.57\%$, and HSCR's AUC is no lower than the strongest baseline in all 25 cells. The corresponding macro-average MTR decreases from $4.80$ for the strongest baseline to $3.87$. Aggregated by node count, HSCR's mean AUC increases from $0.9942$ at $n=50$ to $0.9985$ at $n=200$, with positive mean gains at all five sizes. Aggregated by the number of added edges, mean AUC decreases from $0.9982$ to $0.9950$ as $k$ increases, but exceeds the corresponding strongest baseline at all five settings.
\par\endgroup

\begingroup\emergencystretch=2em
\noindent\textbf{Additional evaluations.} Appendix~\ref{app:data-efficiency} evaluates training-data efficiency under window localization. HSCR exceeds the best baseline at all four tested training fractions, although intermediate performance is not monotonic. Appendix~\ref{app:transfer} examines single-source cross-network transfer under synchronous inference using real temporal topologies with simulated states. Transfer retains predictive discrimination but generally falls below in-domain training, with performance varying by source--target direction.
\par\endgroup

\FloatBarrier
\FixedSpaceSection{Conclusion}

\begingroup\setlength{\parskip}{0pt}
\begingroup\emergencystretch=1em
HSCR ranks connections that have formed but remain unrecorded by organizing node-state evidence through global network context, candidate-local structure, and candidate-set relationships, without prespecified dynamical equations or parameters. Experiments on five temporal-network benchmarks show improved candidate discrimination and true-edge ranks under synchronous, window-aggregated, and structurally delayed observations. Ablations indicate that state evidence benefits from its correct correspondence with candidate-local structure and network positions. On Reddit, ranking ability remains after excluding the direct contribution of contemporaneous edge-addition events from state statistics. These findings support using state responses to complement incomplete structural records within a specified candidate domain, without reconstructing the complete network. In online communities, information networks, and other networked systems, the resulting rankings could guide structural verification and record maintenance, although their operational benefit remains to be evaluated.
\par\endgroup
\par\endgroup

\bibliographystyle{ACM-Reference-Format}
\bibliography{refs}

\appendix
\section{Supplementary Method Details}
\label{app:method}

This appendix gives implementation-level definitions omitted from the main text. It uses the same notation and parameters as Section~\ref{sec:method}; no candidate nonedge is inserted into the visible graph.

\subsection{Complete Observable Evidence}
\label{app:evidence}
Table~\ref{tab:hscr-edge-evidence} lists the raw edge-level evidence. The nine topology features are endpoint degrees $d_u,d_v$, their sum and product, common-neighbor count, Jaccard coefficient, Adamic--Adar index, Resource Allocation index, and visible-graph shortest-path distance. The ten state features comprise $x^s$, $x^e$, $\Delta x$, and $|\Delta x|$ for both endpoints, together with endpoint differences at the initial and final times. Endpoint degrees are ordered as $[\min(d_u,d_v),\max(d_u,d_v)]$; the remaining topology statistics and state differences are symmetric. Applying $\rho$ to the four endpoint-wise state quantities yields $9+14=23$ dimensions. Canonicalization guarantees endpoint-exchange consistency, but not invariance to arbitrary node relabeling.
\begin{table}[H]
 \centering
 \footnotesize
 \renewcommand{\arraystretch}{0.78}
 \setlength{\tabcolsep}{3pt}
 \caption{Raw observable edge-level evidence used by HSCR.}
 \label{tab:hscr-edge-evidence}
 \begin{tabularx}{\linewidth}{ll>{\raggedright\arraybackslash}X}
  \toprule Evidence & Dim. & Features \\ \midrule
  Topology & 9 & Endpoint degrees, degree sum and product, common neighbors, Jaccard, Adamic--Adar, Resource Allocation, and visible-graph shortest-path distance. \\ 
  State & 10 & $x^s$, $x^e$, $\Delta x$, and $|\Delta x|$ for both endpoints, plus endpoint state differences at initial and final times. \\ \bottomrule
 \end{tabularx}
\end{table}

The visible-graph encoder starts with $h_i^{(0)}=\operatorname{MLP}_{\mathrm{node}}(a_i)$ and applies $L_g$ mean-message-passing layers over $E_{\mathrm{obs}}$:
\begin{align}
 \bar h_i^{(\ell)}&=|\mathcal N_i|^{-1}\sum_{j\in\mathcal N_i}h_j^{(\ell)},\nonumber\\
 h_i^{(\ell+1)}&=\operatorname{LN}\!\left(h_i^{(\ell)}+\operatorname{Dropout}\left[\operatorname{ReLU}\left(W_s^{(\ell)}h_i^{(\ell)}+W_n^{(\ell)}\bar h_i^{(\ell)}\right)\right]\right),
 \label{eq:hscr-visible-message-passing}
\end{align}
\begingroup
\predisplaypenalty=0
where the neighbor mean is zero for an isolated node. State refinement is
\begin{align}
 r_i&=\operatorname{Enc}_{\mathrm{state}}([x_i^s,x_i^e,\Delta x_i]),\nonumber\\
 \delta_i^s&=\operatorname{MLP}_{\mathrm{ref}}([h_i^g,r_i,|h_i^g-r_i|,h_i^g\odot r_i]),\nonumber\\
 \gamma_i^s&=\tanh(\operatorname{Gate}_{\mathrm{state}}([h_i^g,r_i])),\nonumber\\
 h_i&=h_i^g+\gamma_i^s\odot\delta_i^s,
 \label{eq:hscr-state-refinement}
\end{align}
\endgroup
with a zero-initialized final gate. The endpoint interaction and two explicit-evidence branches are
\begin{align}
 \psi_e&=[h_{u_e},h_{v_e},|h_{u_e}-h_{v_e}|,h_{u_e}\odot h_{v_e}],\label{eq:hscr-endpoint-interaction}\\
 r_e^{\mathrm{top}}&=\operatorname{Enc}_{\mathrm{top}}([\psi_e,f_e^{\mathrm{top}}]),\quad r_e^{\mathrm{state}}=\operatorname{Enc}_{\mathrm{state-edge}}([\psi_e,f_e^{\mathrm{state}}]).\label{eq:hscr-edge-branches}
\end{align}

\subsection{Local Structural Encoder Details}
\label{app:local}
If $\mathcal V_e^{\mathrm{loc}}$ exceeds $M$, nodes are retained using nearest-endpoint distance, summed endpoint distances, degree, and node index as deterministic tie breakers. For local node $j$, let $r_{j|e}^{\mathrm{near}}$ and $r_{j|e}^{\mathrm{far}}$ be its smaller and larger endpoint distances, and let $r_{j|e}$ denote its symmetric role (endpoint, common neighbor, one-sided branch, balanced, or other). The initial token is
\begin{equation}
 \begin{aligned}
 \ell_{j|e}^{(0)}=\operatorname{LN}\bigl(&h_j+W_dd_j+\operatorname{Emb}_{\mathrm{near}}(r_{j|e}^{\mathrm{near}})\\
 &+\operatorname{Emb}_{\mathrm{far}}(r_{j|e}^{\mathrm{far}})+\operatorname{Emb}_{\mathrm{role}}(r_{j|e})\bigr).
 \end{aligned}
 \label{eq:hscr-local-node-token}
\end{equation}
With adjacency/self mask $\mathcal A_e(j,k)$, local attention uses
\begin{align}
 a_{jk}^{(\ell,e)}&=\frac{(W_Q\ell_{j|e}^{(\ell)})^\top W_K\ell_{k|e}^{(\ell)}}{\sqrt d},\nonumber\\
 \omega_{jk}^{(\ell,e)}&=\operatorname{softmax}_{k:\,\mathcal A_e(j,k)=1}(a_{jk}^{(\ell,e)}),\nonumber\\
 \ell_{j|e}^{(\ell+1)}&=\operatorname{LN}\left(\ell_{j|e}^{(\ell)}+\operatorname{Dropout}\left[W_O\sum_k\omega_{jk}^{(\ell,e)}W_V\ell_{k|e}^{(\ell)}\right]\right).
 \label{eq:hscr-local-attention}
\end{align}
After $L_\ell$ layers, endpoint pooling is
\begin{align}
 \bar\ell_e&=(\widetilde\ell_{u_e|e}+\widetilde\ell_{v_e|e})/2,\quad q_e^{\mathrm{loc}}=W_Q^{\mathrm{loc}}\bar\ell_e,\nonumber\\
 \beta_{j|e}^{\mathrm{loc}}&=\operatorname{softmax}_j((q_e^{\mathrm{loc}})^\top W_K^{\mathrm{loc}}\widetilde\ell_{j|e}),\nonumber\\
 p_e^{\mathrm{loc}}&=\sum_j\beta_{j|e}^{\mathrm{loc}}W_V^{\mathrm{loc}}\widetilde\ell_{j|e}.
 \label{eq:hscr-local-pooling}
\end{align}
The local increment is
\begin{align}
 u_e^{\mathrm{loc}}&=\operatorname{MLP}_{\mathrm{loc}}([\bar\ell_e,p_e^{\mathrm{loc}},|\bar\ell_e-p_e^{\mathrm{loc}}|,\bar\ell_e\odot p_e^{\mathrm{loc}}]),\nonumber\\
 c_e^{\mathrm{loc}}&=c_e+\gamma_e^{\mathrm{loc}}\odot u_e^{\mathrm{loc}},
 \label{eq:hscr-local-update-detail}
\end{align}
where $\gamma_e^{\mathrm{loc}}=\tanh(\operatorname{Gate}_{\mathrm{loc}}([c_e,u_e^{\mathrm{loc}}]))$.

\subsection{Candidate-Set Context and Complexity}
\label{app:set}
\begingroup
\emergencystretch=8em
The complete-set summary in the main text is additive over candidates. For a disjoint partition $\mathcal C=\bigcup_b\mathcal C_b$, $N_k=\sum_bN_k^{(b)}$ and $D_k=\sum_bD_k^{(b)}$, enabling blockwise accumulation without changing $\mathcal C$. Each candidate reads these summaries as follows:
\par\endgroup
\begingroup
\allowdisplaybreaks[4]
\begin{align}
 \beta_{ek}^{\mathrm{set}}&=\operatorname{softmax}_k\left((W_Qc_e^{\mathrm{loc}})^\top W_KH_k/\sqrt d\right),\nonumber\\
 u_e^{\mathrm{set}}&=\sum_{k=1}^K\beta_{ek}^{\mathrm{set}}W_VH_k,\nonumber\\
 \delta_e^{\mathrm{set}}&=\operatorname{MLP}_{\mathrm{set}}([c_e^{\mathrm{loc}},u_e^{\mathrm{set}}]),\nonumber\\
 \gamma_e^{\mathrm{set}}&=\tanh(\operatorname{Gate}_{\mathrm{set}}([c_e^{\mathrm{loc}},u_e^{\mathrm{set}}])),\nonumber\\
 z_e&=c_e^{\mathrm{loc}}+\gamma_e^{\mathrm{set}}\odot\delta_e^{\mathrm{set}}.
 \label{eq:hscr-set-readback}
\end{align}
\endgroup
The summary is permutation-invariant; readback and scores reorder with candidates, so scoring is permutation-equivariant. The set-interaction cost is $O(mK)$, excluding node, local, and scoring modules.

\subsection{Training Objective and Procedure}
\label{app:training}
Let $\mathcal P$ be positive indices, $\mathcal N_s$ sampled negatives, and $\mathcal Q$ positive--negative pairs. The three losses are
\begin{align}
 \mathcal L_{\mathrm{pair}}&=|\mathcal Q|^{-1}\sum_{(p,n)\in\mathcal Q}\operatorname{softplus}(-(s_p-s_n)),\label{eq:hscr-pairwise-loss}\\
 \mathcal L_{\mathrm{set}}&=\max\left(0,1-|\mathcal P|^{-1}\sum_{p\in\mathcal P}s_p+|\mathcal N_s|^{-1}\sum_{n\in\mathcal N_s}s_n\right),\label{eq:hscr-set-loss}\\
 \widehat q&=\operatorname{CountHead}\left(K^{-1}\sum_kH_k\right),\nonumber\\
 \mathcal L_{\mathrm{count}}&=\left(\widehat q-\log(1+|\mathcal P|)\right)^2.
 \label{eq:hscr-count-loss}
\end{align}
The count head is auxiliary and does not enter $s_e$. Training first learns the base node encoder, edge representations, and scoring head. From the validation-selected base, state, local, and set modules are activated; the base is initially frozen and then jointly fine-tuned at a lower learning rate. Validation selects parameters, while testing uses the complete candidate domain. All incremental gates are zero-initialized, giving
\begin{equation}
 h_i=h_i^g,\qquad c_e^{\mathrm{loc}}=c_e,\qquad z_e=c_e
 \label{eq:hscr-zero-gate-fallback}
\end{equation}
at the start of stage two. Thus the added modules begin with zero increments while preserving the base ranking.

\section{Observation Scenarios}
\label{app:observation-scenarios}

\begin{figure*}[!t]
  \centering
  \includegraphics[width=0.9\linewidth]{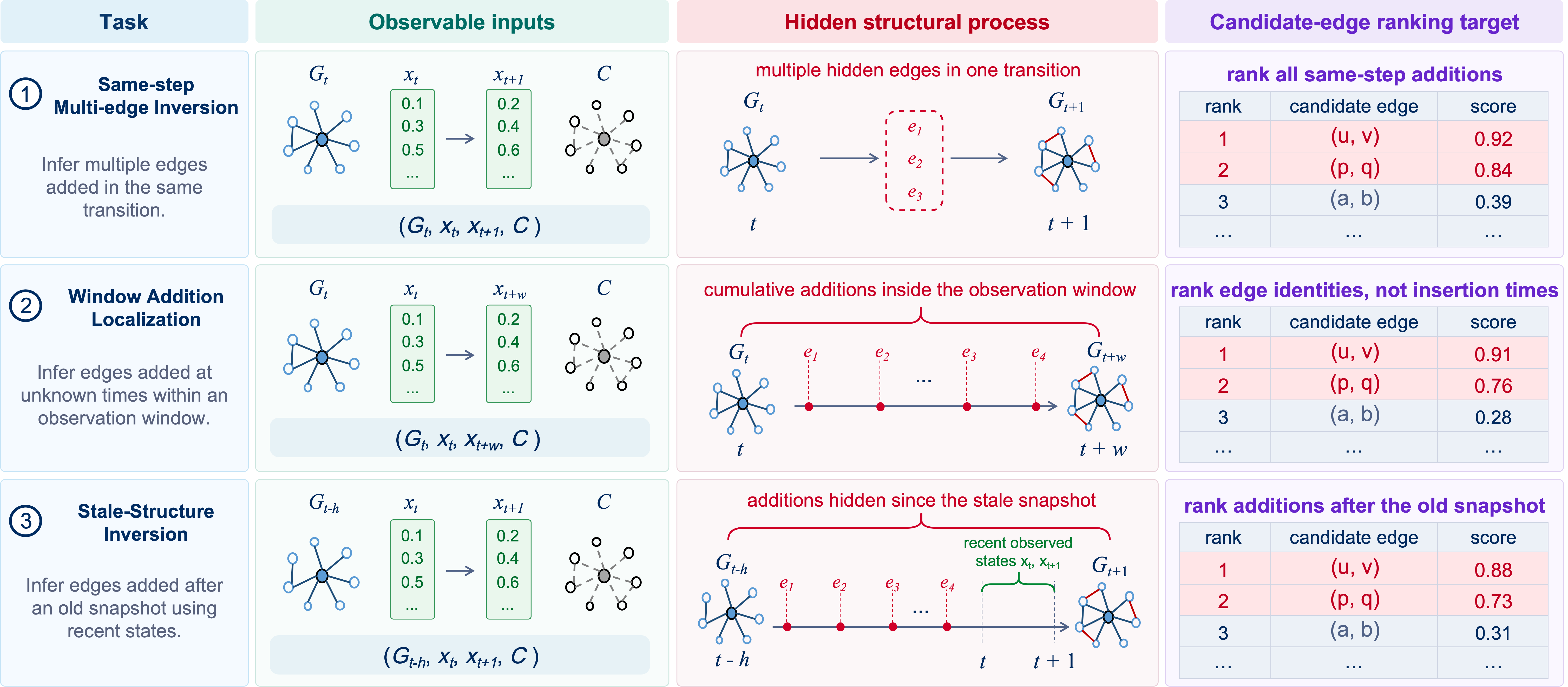}
  \caption{Temporal alignments for hidden edge-addition inference: synchronous, window-based, and stale-snapshot settings. Each ranks candidate edges without estimating insertion times.}
  \label{fig:observation-scenarios}
\end{figure*}

Figure~\ref{fig:observation-scenarios} summarizes the temporal correspondence among the visible structure, state observations, and added-edge targets.

\setkeys{acmart.cls}{balance=false}
\nobalance
\section{Additional Experimental Details and Results}
\label{app:experiments}

\subsection{Reddit Dataset Construction}
\label{app:reddit-data}

\begingroup
\emergencystretch=2em

The Reddit benchmark is derived from the SNAP subreddit hyperlink title-event collection~\cite{leskovec2014snap}. We scan 571,927 valid records covering 54,075 subreddits from December 2013 through April 2017, retain the 300 subreddits with the highest incident-event counts, discard other events and self-loops, and canonicalize pairs as undirected edges, leaving 149,897 retained events. Calendar-month aggregation yields 41 bins and cumulative graphs $G_t=(V,E_t)$; for adjacent bins, the target set is
\begin{equation}
 \Delta E_t=E_{t+1}\setminus E_t.
 \label{eq:reddit-target-edges}
\end{equation}
We retain 40 nonzero windows satisfying $2\leq|\Delta E_t|\leq2000$, where each target edge denotes the first observed pair occurrence. Chronological splitting gives 24 training, 7 validation, and 7 test windows with a one-window temporal buffer.

For each subreddit and month, we aggregate incident hyperlink activity, mean link sentiment, and the first numeric text attribute recorded for each event. Let $a$, $s$, and $p$ denote these quantities; after independently z-scoring the three components, the raw state is
\begin{equation}
 x^{\mathrm{raw}}=z(\log(1+a))+0.5\,z(s)+0.1\,z(\log(1+\max(p,0))).
 \label{eq:reddit-state-construction}
\end{equation}
The state matrix is z-scored once more over all node--month entries. For each window, candidates combine current-graph nonedges within three hops with a global 25\% sample capped at 2,000; each target is paired with five degree-matched negatives using tolerance two. Target edges provide labels only and do not enter the observed graph or state features. Each window contains 31,456 candidates and 412 target edges on average.

The \emph{full-state} view uses all retained events in each month. The \emph{existing-link activity-state} view excludes, from node-state aggregation only, events on edges first appearing in the target month; the structural snapshots, candidate sets, and target labels remain unchanged. Thus, target-month first-appearance events still define the structural targets and evaluation labels, but do not contribute directly to the corresponding state statistics. This comparison tests whether ranking can rely on existing structure and activity rather than the direct contribution of target-month additions. It is a partial-observation evaluation constructed from public records, not evidence that genuinely missing platform logs have been recovered. All derived samples are generated deterministically under the preprocessing, state-construction, candidate-generation, and temporal-splitting rules specified above, and all conclusions are restricted to this candidate-ranking setting.
\par\endgroup

\subsection{Dynamical Rules and Default Parameters}
\label{app:dynamics}
Table~\ref{tab:dynamics_registry} lists the state-update rules and default parameters. Here, $\bar{x}_{\mathcal N(i)}$ denotes the neighbor mean and $\sigma(\cdot)$ the sigmoid.

\begin{table*}[!t]
  \centering
  \caption{Dynamical rules and default parameters used to generate state observations.}
  \label{tab:dynamics_registry}
  \small
  \setlength{\tabcolsep}{5pt}
  \begin{tabular}{lll}
    \toprule
    Dynamics & State-update mechanism & Main parameters \\
    \midrule
    DeGroot & $x_i' = \alpha x_i+(1-\alpha)\bar{x}_{\mathcal N(i)}$ & $\alpha=0.5$ \\
    Nonlinear response & $x_i'=\sigma\!\left(\beta[\alpha x_i+(1-\alpha)\bar{x}_{\mathcal N(i)}-\theta]\right)$ & $\alpha=0.5,\ \beta=5.0,\ \theta=0.5$ \\
    Threshold propagation & $x_i'=(1-\rho)x_i+\rho\,\sigma\!\left(\beta[\bar{x}_{\mathcal N(i)}-\theta]\right)$ & $\theta=0.35,\ \beta=12.0,\ \rho=0.6$ \\
    SIS mean-field & $x_i'=(1-\gamma)x_i+(1-x_i)[1-\exp(-\lambda\bar{x}_{\mathcal N(i)})]$ & $\lambda=0.4,\ \gamma=0.2$ \\
    Coupled logistic & $x_i'=(1-\kappa)r x_i(1-x_i)+\kappa r\bar{x}_{\mathcal N(i)}(1-\bar{x}_{\mathcal N(i)})$ & $r=3.6,\ \kappa=0.3$ \\
    \bottomrule
  \end{tabular}
\end{table*}

\subsection{Data Efficiency}\label{app:data-efficiency}
\begin{figure}[!t]
  \setlength{\abovecaptionskip}{1pt}
  \centering
  \includegraphics[width=0.855\columnwidth,trim=0bp 4.36bp 0bp 0bp,clip]{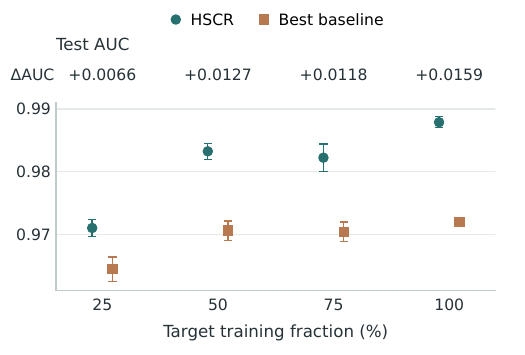}
  \caption{Window-localization data efficiency: mean test AUC $\pm$ SD over three runs.}
  \label{fig:data-efficiency}
\end{figure}

Under window localization, we train on four fractions (25\%, 50\%, 75\%, 100\%) with three runs each, sampling complete trajectories from the six-network training/validation pool and evaluating on the five reported networks. Achieved fractions are 25.5\%, 58.7\%, 80.5\%, and 100\%. Fig.~\ref{fig:data-efficiency} shows HSCR above the best baseline at all settings (mean differences $0.0066$--$0.0159$), with AUC $0.9710$ at the smallest fraction and $0.9879$ at full data; intermediate performance is not monotonic.

\subsection{Cross-Network Transfer}\label{app:transfer}
\begin{figure}[!t]
  \setlength{\abovecaptionskip}{1pt}
  \centering
  \includegraphics[width=0.95\columnwidth]{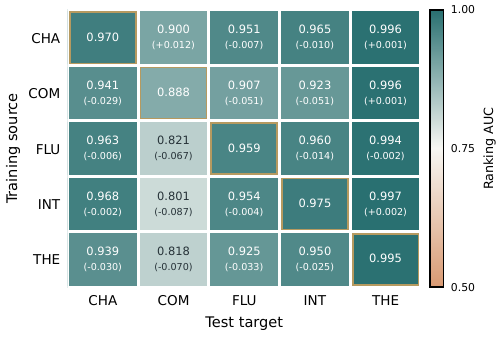}
  \caption{Single-source cross-network transfer AUC; rows and columns denote training sources and test targets, parentheses show changes from target in-domain AUC, and gold borders mark in-domain references.}
  \label{fig:single-source-transfer}
\end{figure}

\newcommand{\TransferMeanAUC}{0.9335}
\newcommand{\TransferAbove}{17}
\newcommand{\TransferCHAWithin}{0.9698}
\newcommand{\TransferCHAMean}{0.9529}
\newcommand{\TransferCOMWithin}{0.8879}
\newcommand{\TransferCOMMean}{0.8351}
\newcommand{\TransferFLUWithin}{0.9586}
\newcommand{\TransferFLUMean}{0.9346}
\newcommand{\TransferINTWithin}{0.9747}
\newcommand{\TransferINTMean}{0.9495}
\newcommand{\TransferTHEWithin}{0.9953}
\newcommand{\TransferTHEMean}{0.9956}
\newcommand{\TransferCHAToCOM}{0.9000}
\newcommand{\TransferCHAToFLU}{0.9512}
\newcommand{\TransferCHAToINT}{0.9647}
\newcommand{\TransferCHAToTHE}{0.9960}
\newcommand{\TransferCOMToCHA}{0.9409}
\newcommand{\TransferCOMToFLU}{0.9074}
\newcommand{\TransferCOMToINT}{0.9233}
\newcommand{\TransferCOMToTHE}{0.9958}
\newcommand{\TransferFLUToCHA}{0.9633}
\newcommand{\TransferFLUToCOM}{0.8211}
\newcommand{\TransferFLUToINT}{0.9603}
\newcommand{\TransferFLUToTHE}{0.9936}
\newcommand{\TransferINTToCHA}{0.9682}
\newcommand{\TransferINTToCOM}{0.8013}
\newcommand{\TransferINTToFLU}{0.9542}
\newcommand{\TransferINTToTHE}{0.9969}
\newcommand{\TransferTHEToCHA}{0.9394}
\newcommand{\TransferTHEToCOM}{0.8180}
\newcommand{\TransferTHEToFLU}{0.9255}
\newcommand{\TransferTHEToINT}{0.9496}

We test single-source transfer under synchronous inference using real temporal topologies with simulated states. HSCR is trained on each of five networks and tested on the other four (20 directed pairs); targets are excluded from training and model selection, and transfer/in-domain references use identical target instances and candidates.

\begingroup\setlength{\parskip}{0pt}
\begingroup\emergencystretch=2em
The 20 pairs have mean AUC \TransferMeanAUC{}, with \TransferAbove{} above $0.90$ (Fig.~\ref{fig:single-source-transfer}). Transfer varies by direction: Complex Networks$\rightarrow$\allowbreak Thermodynamics gives \TransferCOMToTHE{}, versus \TransferTHEToCOM{} in reverse. Four of five targets have lower transfer than in-domain performance; Complex Networks is an example (\TransferCOMMean{} vs. \TransferCOMWithin{}), while Thermodynamics is slightly higher (\TransferTHEMean{} vs. \TransferTHEWithin{}). Thus HSCR retains identification ability without target training, but transfer does not replace in-domain learning for most targets.
\par\endgroup
\par\endgroup

\end{document}